\documentstyle[preprint,pre,aps]{revtex}

\tightenlines
\begin{document}
\draft

\title{Critical behavior of period doublings in
       coupled inverted pendulums
      }
\author{Sang-Yoon Kim${}^{a,b}$
\footnote{Electronic address: sykim@cc.kangwon.ac.kr}
 and Bambi Hu${}^{b,c}$
       }
 \address{
 ${}^a$Department of Physics, Kangwon National University,
 Chunchon, Kangwon-Do 200-701, Korea~
 ${}^b$ Centre for Nonlinear Studies and Department of Physics,
 Hong Kong Baptist University, Hong Kong, China ~
 ${}^c$ Department of Physics, University of Houston, Houston,
 TX 77204
 }

\maketitle

\begin{abstract}
We study the critical behaviors of period doublings in $N$
$(N=2,3,4,\dots)$ coupled inverted pendulums by varying the driving
amplitude $A$ and the coupling strength $c$. It is found that the
critical behaviors depend on the range of coupling interaction. In
the extreme long-range case of global coupling, in which each
inverted pendulum is coupled to all the other ones with equal
strength, the zero-coupling critical point and an infinity of
critical line segments constitute the same critical set in the
$A-c$ plane, independently of $N$. However, for any other
nonglobal-coupling cases of shorter-range couplings,
the structure of the critical set becomes different from that for
the global-coupling case, because of a significant change in the
stability diagram of periodic orbits born via period doublings.
The critical scaling behaviors on the critical set are also found
to be the same as those for the abstract system of the coupled
one-dimensional maps.
\end{abstract}


%
%

\narrowtext

\section{Introduction}
\label{sec:Int}

The nonlinear dynamics of coupled nonlinear oscillators has
attracted considerable attention in recent years. Such coupled
oscillators are used to model many physical, chemical, and
biological systems such as coupled p-n junctions \cite{Jeffries},
Josephson-junction arrays \cite{Hadley}, the charge-density waves
\cite{Strogatz}, chemical-reaction systems \cite{Kuramoto}, and
biological-oscillation systems \cite{Winfree}. They exhibit
diverse bifurcations, multistability, chaos, pattern formation,
and so on.

The coupled nonlinear oscillators studied here are coupled
inverted pendulums, consisting of $N$ identical inverted pendulum
coupled through some interaction mechanism. We first
consider a constituent element of the coupled dynamical
system, i.e., a single parametrically forced pendulum with
a vertically-oscillating suspension point. It can be described by
a normalized equation of motion \cite{PFP},
\begin{equation}
 {\ddot x}= f(x,\dot{x},t)
          = - 2 \pi \beta \Omega {\dot  x}
            - 2 \pi (\Omega^2 - A \cos 2 \pi t) \sin 2 \pi x,
\label{eq:IP}
\end{equation}
where $x$ is a normalized angle with range $x \in [0,1)$,
$\beta$ is a normalized damping parameter, $\Omega$ is the normalized
natural frequency of the unforced pendulum, and $A$ is the normalized
driving amplitude of the vertical oscillation of the suspension
point, respectively. This parametrically forced pendulum has an
``inverted'' stationary state, corresponding to the vertically-up
configuration with $x={1 \over 2}$. It is well known that as the
parameter $A$ is increased, the inverted pendulum undergoes a
cascade of ``resurrections,'' i.e., it becomes stabilized after
its instability, destabilize again and so forth {\it ad infinitum}
\cite{Corben,Kapitza,Stoker,Levi}. Recently, we have studied
bifurcations and transitions to chaos associated with such
resurrections of the inverted pendulum \cite{Kim1}. For each case
of the resurrections, the stabilized inverted state exhibits an
infinite sequence of period-doubling bifurcations accumulating at a
period-doubling transition point $A^*$, beyond which chaos sets in.
Consequently, an infinite series of period-doubling transitions to
chaos occur successively with increasing $A$.
This is in contrast to the one-dimensional (1D) map
\cite{Feigenbaum}, where only single period-doubling transition
to chaos takes place. However, the critical scaling behaviors at
each $i$th period-doubling transition point $A^*_i$
$(i=1,2,3, \dots)$ are the same as those for the 1D map.

In this paper we study the critical behaviors of period doublings
in the system of $N$ $(N=2,3,4,\dots)$ symmetrically coupled
inverted pendulums by varying the driving amplitude $A$ and the
strength $c$ of coupling between the inverted pendulums, and also
compare them with those for the abstract system of the coupled
1D maps \cite{KK,Kim2}. The ``coupling effect'' of the strength
and range of coupling on the critical behaviors are
particularly investigated. Both the structure of the critical set
and the critical scaling behaviors for the coupled inverted
pendulums are found to be the same as those for the coupled 1D maps
found by one of us (Kim) and Kook \cite{Kim2}.

This paper is organized as follows. We first introduce $N$
symmetrically coupled inverted pendulums in Sec.~\ref{sec:CIP} and
discuss their dynamical symmetries and couplings. Bifurcations
associated with stability of periodic orbits and Lyapunov
exponents in the coupled inverted pendulums are also discussed in
Sec.~\ref{sec:SBL}. We then investigate the critical behaviors of
period doublings in the coupled inverted pendulums in
Sec.~\ref{sec:CSB}. As in the single inverted pendulum \cite{Kim1},
the coupled inverted pendulums undergo multiple period-doubling 
transitions to chaos [e.g., see Figs.~\ref{SD2}(a), \ref{SD3}(a) 
and \ref{SD3}(b) for the ``stability trees'' associated with the 
first, second, and third period-doubling transitions to chaos, 
respectively]. For each period-doubling transition to chaos, the 
critical behaviors vary depending on whether or not the coupling is 
global. In the extreme long-range case of global coupling, the 
zero-coupling critical point with $c=0$ and an infinity of critical 
line segments lying on the line $A=A^*_i$ constitute the same 
critical set in the $A-c$ plane, irrespectively of $N$. However, 
for any other nonglobal-coupling cases of shorter-range couplings, 
a significant change occurs in the stability diagram of 
$2^n$-periodic $(n=0,1,2, \dots)$ orbits in the $A-c$ plane, and 
consequently the structure of the critical set becomes different 
from that for the global-coupling case. It is also found that the 
critical scaling behaviors on the critical set are the same as 
those for the abstract system of the coupled 1D maps \cite{Kim2}. 
Finally, a summary is given in Sec.~\ref{sec:Sum}.

\section{Symmetries and Couplings in The Coupled Inverted Pendulums}
\label{sec:CIP}

In this section we introduce $N$ symmetrically coupled inverted
pendulums and then discuss their symmetries and couplings.

Consider $N$ symmetrically coupled inverted pendulums with a periodic
boundary condition,
\begin{equation}
{\ddot{x}}_m = f(x_m,{\dot x}_m,t) + g(x_m,x_{m+1},\dots,x_{m-1}),
\;\;m=1,2,\dots,N.
\label{eq:MCIP1}
\end{equation}
Here the periodic boundary condition imposes $x_m(t) = x_{m+N}(t)$ for
all $m$, the function $f(x,{\dot x},t)$ is given in Eq.~(\ref{eq:IP}),
and $g(x_1,\dots,x_N)$ is a coupling function, obeying the condition
\begin{equation}
g(x,\dots,x) = 0 \;{\rm for\;all\;}x.
\label{eq:CC}
\end{equation}

The second-order differential equations (\ref{eq:MCIP1}) are reduced
to a set of first-order differential equations,
\begin{mathletters}
\begin{eqnarray}
{\dot x_m} &=& y_m, \\
{\dot y_m} &=& f(x_m,y_m,t) + g(x_m,x_{m+1},\dots,x_{m-1}),
\;\;m=1,2,\dots,N.
\end{eqnarray}
\label{eq:MCIP2}
\end{mathletters}
Consider an initial orbit point ${\bf{z}}(0)$
$[=(z_1(0),\dots,z_N(0))]$, where $z_i=(x_i,y_i)$ $(i=1,\dots,N)$.
Then its Poincar\'{e} maps can be computed by sampling the orbit
points ${\bf{z}}(m)$ at the discrete time $t=m$ $(m=1,2,3,\dots)$.
We will call the transformation ${\bf{z}}(m) \rightarrow
{\bf{z}}(m+1)$ the Poincar\'{e} map and write ${\bf{z}}(m+1)=
P({\bf{z}}(m))$.

The $2N$-dimensional Poincar\'{e} map $P$ has a cyclic permutation
symmetry such that
\begin{equation}
\sigma^{-1} P \sigma ( {\bf z}) = P({\bf z}) \;\;{\rm for \;all\;}z,
\label{eq:CPS}
\end{equation}
where $\sigma$ is a cyclic permutation of ${\bf z}$ such that
$\sigma (z_1,z_2,\dots,z_N) = (z_2,\dots,z_N,z_1)$ and $\sigma^{-1}$
is its inverse. The set of all fixed points of $\sigma$ forms a
two-dimensional (2D) synchronization plane, on which
\begin{equation}
x_1 = \cdots = x_N, \,\, y_1 = \cdots = y_N.
\end{equation}
It follows from Eq.~(\ref{eq:CPS}) that the cyclic permutation
$\sigma$ commutes with the Poincar\'{e} map $P$, i.e.,
$\sigma P = P \sigma$. Consequently, the 2D synchronization plane
becomes invariant under $P$, i.e., if a point ${\bf z}$ lies on the
2D synchronization plane, then its image $P({\bf z})$ also lies on it.
An orbit is called a(n) (in-phase) synchronous orbit if it lies on the
2D invariant synchronization plane, i,e, it satisfies
\begin{equation}
x_1 (t) = \cdots = x_N (t) \equiv x^*(t), \,\,
y_1 (t) = \cdots = y_N (t) \equiv y^*(t).
\label{eq:SO}
\end{equation}
Otherwise it is called an (out-of-phase) asynchronous orbit.
Here we study only the synchronous orbits. They can be easily found from
the uncoupled inverted pendulum (\ref{eq:IP}), because the coupling
function $g$ satisfies the condition (\ref{eq:CC}). Note also that for
these synchronous orbits, the Poincar\'{e} map $P$ also has the
inversion symmetry such that
\begin{equation}
S P S ( {\bf z}) = P({\bf z}) \;\;{\rm for \;all\;}z,
\label{eq:IS}
\end{equation}
where $S({\bf z})=-{\bf z}$.
If a synchronous orbit $\{ {\bf z}(t) \}$ of $P$ is invariant under $S$,
it is called a symmetric orbit. Otherwise, it is called an
asymmetric orbit and has its ``conjugate'' orbit $S\{ {\bf z}(t) \}$.

We now discuss the couplings between the inverted pendulums. Consider
an element, say the $m$th element, in the $N$ coupled inverted
pendulums. Then the $(m\pm\delta)$th elements are called the
$\delta$th neighbors of the $m$th element. Here we consider the case
where the coupling extends to the
$K$th $[1 \leq K \leq {N \over 2} ({ {N-1} \over 2})$ for
even (odd) $N$] neighbor(s) with equal strength. Hereafter we will
call the number $K$ the range of the coupling interaction.

A general form of coupling for odd $N$ $(N \geq 3)$ is given by
\begin{eqnarray}
g(x_1,\dots,x_N) &=& {\frac {c} {2K+1}}
  {\sum_{l=-K}^{K}} [u(x_{1+l}) - u(x_1)] \nonumber \\
&=& c \left[ {\frac {1}{2K+1}} {\sum_{l=-K}^{K}} u(x_{1+l}) - u(x_1)
\right], \nonumber \\
&&K=1,\dots,{\frac {N-1} {2}},
\label{eq:GCF}
\end{eqnarray}
where $c$ is a coupling parameter and $u$ is a function of one
variable. Note that the coupling extends to the $K$th neighbors with
equal coupling strength, and the function $g$ satisfies the condition
(\ref{eq:CC}).
The extreme long-range interaction for
$K= {{\frac {N-1} {2}}}$ is
called a global coupling, for  which  the coupling  function $g$
becomes
\begin{eqnarray}
g(x_1,\dots,x_N)  &=&  {\frac  {c}  {N}}{\sum_{m=1}^{N}}  [u(x_{m}) -
u(x_1)] \nonumber \\
&=& c  \left[ {\frac  {1}{N}} {\sum_{m=1}^{N}} u(x_m)  - u(x_1)
\right].
\label{eq:GC}
\end{eqnarray}
This is a kind of mean-field coupling, in which each inverted pendulum is
coupled to all the  other ones with equal coupling strength. All
the other couplings with $K < {{\frac {N-1} {2}}}$ (e.g.,
nearest-neighbor coupling with $K=1$) will be referred to as nonglobal
couplings. The $K=1$ case for $N=3$ corresponds to both the global
coupling and the nearest-neighbor coupling.

We next consider the case of even $N$ $(N \geq 2)$.
The  form of  coupling of  Eq.~(\ref{eq:GCF}) holds  for the  cases of
nonglobal couplings with $K=1,\dots,{{\frac{N-2}{2}}}$ $(N \geq 4)$.
The global coupling for $K= {{\frac {N} {2}}}$ $(N \geq 2)$ also
has the form of Eq.~(\ref{eq:GC}), but it cannot  have the form of
Eq.~(\ref{eq:GCF}), because there exists only one farthest neighbor
for $K= { {\frac{N}{2}} }$, unlike the case of odd $N$.
The  $K=1$ case  for $N=2$  also  corresponds to  the nearest-neighbor
coupling as well as to the global coupling, like the $N=3$ case.

\section{Stability, Bifurcations, and Lyapunov Exponents of
         Synchronous Orbits}
\label{sec:SBL}

In this section we first discuss stability of synchronous periodic
orbits in the Poincar\'{e} map $P$ of the coupled inverted pendulums,
using the Floquet theory \cite{Lef1}. Bifurcations associated with the
stability and Lyapunov exponents are then discussed.

The stability analysis of an orbit in many-coupled inverted pendulums
can be conveniently carried out by Fourier-transforming with respect
to the discrete space $\{m\}$ \cite{Kapral}. Consider an orbit
$\{ {x_m}(t)\;  ; \;m=1,\dots,N \}$ of the $N$ coupled inverted
pendulums (\ref{eq:MCIP1}). The discrete spatial Fourier transform of
the orbit is:
\begin{eqnarray}
{\cal F}[{x_m(t)}] &\equiv& {\frac{1}{N}} {\sum_{m=1}^{N}}
{e^{-2{\pi}imj/N}} {x_m}(t) = {\xi}_j(t), \nonumber \\
&&\;\;\;\;\;\;\;\;\;\; j=0,1,\dots,N-1.
\label{eq:FT}
\end{eqnarray}
The Fourier transform $\xi_j(t)$ satisfies $\xi_j^*(t) = \xi_{N-j}(t)$
($*$ denotes complex conjugate), and
the wavelength of a mode with index $j$ is ${\frac {N}{j}}$ for
$j \leq {{ {\frac {N} {2}}}}$ and
${\frac {N} {N-j}}$ for $j > {{\frac {N} {2}}}$. Here $\xi_0$
corresponds to the synchronous (Fourier) mode of the orbit, while
all the other $\xi_j$'s with nonzero indices $j$ correspond to the
asynchronous (Fourier) modes.

To determine the stability of a synchronous $q$-periodic orbit
[$x_1(t) = \cdots =x_N(t) \equiv x^{*}(t)$ for all $t$ and
$x^*(t) = x^*(t+q)$],
we consider an infinitesimal perturbation $\{ {\delta}x_m(t) \}$
to the synchronous orbit, i.e.,
$x_m(t)=x^{*}(t)+{\delta}x_m(t)$ for $m=1,\dots,N$.
Linearizing the $N$-coupled inverted pendulums (\ref{eq:MCIP1}) at the
synchronous orbit, we obtain:
\begin{eqnarray}
{\delta}{\ddot x_m} &=& { {\partial f(x^*,\dot{x}^*,t)} \over
{\partial x^*}}
{\delta}x_m + { {\partial f(x^*,\dot{x}^*,t)} \over {\partial
\dot{x}^*} }
{\delta}{\dot x}_m \nonumber \\
 &&   +   {\sum_{l=1}^{N}}   {G_l}(x^{*})\;   {\delta}x_{l+m-1},
\label{eq:LE}
\end{eqnarray}
where
\begin{equation}
G_l(x) \equiv
\left. { \frac{\partial g(x_1,\dots,x_N)}{\partial x_l} }
\right |_{x_1=\cdots=x_N=x}.
\label{eq:RCF}
\end{equation}
Hereafter the  functions $G_l$'s  will be called  ``reduced'' coupling
functions of $g(x_1,\dots,x_N)$.

Let  ${\delta  {\xi}_j}(t)$  be  the  Fourier  transform  of  {$\delta
x_m(t)$},
i.e.,
\begin{eqnarray}
\delta \xi_j =
{\cal F}[{\delta x_m(t)}] &=& {\frac{1}{N}} {\sum_{m=1}^{N}}
{e^{-2{\pi}imj/N}} {\delta x_m}, \nonumber \\
&&\;\;\;\;\;\;\;\;\;j=0,1,\dots,N-1.
\end{eqnarray}
Here $\delta \xi_0$ is the synchronous-mode perturbation, and all the
other $\delta \xi_j$'s with nonzero indices $j$ are the
asynchronous-mode perturbations.
Then the Fourier transform of Eq.~(\ref{eq:LE}) becomes:
\begin{eqnarray}
\delta {\ddot{\xi}}_j &=&
{{\partial f(x^*,\dot{x}^*,t)} \over {\partial \dot{x}^*}}
{\delta}{\dot{\xi}}_j +
[{{\partial f(x^*,\dot{x}^*,t)} \over {\partial x^*}}  \nonumber \\
&& + \sum_{l=1}^{N} {G_l}(x^{*}) {e^{2 \pi i(l-1)j/N}}]
{\delta {\xi}_j}, \;\; j=0,1,\dots,N-1.
\label{eq:LM1}
\end{eqnarray}
Note that all the modes $\delta \xi_j$'s become decoupled for the
synchronous orbit.

The equation (\ref{eq:LM1}) can also be put into the following form:
\begin{eqnarray}
\left(
\begin{array}{l}
\delta{\dot{\xi}}_j \\
\delta{\dot{\eta}}_j
\end{array}
\right)
&=& J_j(t)
\left(
\begin{array}{l}
\delta{\xi}_j \\
\delta{\eta}_j
\end{array}
\right), \;\;j=0,1,\dots,N-1,
\label{eq:LM2}
\end{eqnarray}
where
\begin{equation}
J_j(t)=
\left( \begin{array}{cc}
0 & \;\; 1 \\
{{{\partial f(x^*,\dot{x}^*,t)} \over {\partial x^*} } +
{\displaystyle{\sum_{l=1}^{N}}} {G_l}(x^{*}) {e^{2 \pi i(l-1)j/N}}}
&\;\; {{\partial f(x^*,\dot{x}^*,t)} \over {\partial \dot{x}^*} }
    \end{array}
\right).
\label{eq:JML}
\end{equation}
Note that each $J_j$ is a $q$-periodic matrix, i.e.,
$J_j(t) = J_j(t+q)$. Using the Floquet theory \cite{Lef1}, we
study the stability of the synchronous $q$-periodic orbit against the
$j$th-mode perturbation as follows.
Let $\Phi_j(t)=(\phi^{(1)}_j(t),\phi^{(2)}_j(t))$ be a fundamental
solution matrix with $\Phi_j(0) = I$. Here $\phi^{(1)}_j(t)$ and
$\phi^{(2)}_j(t)$ are two independent solutions expressed in column
vector forms, and $I$ is
the $2 \times 2$ unit matrix. Then a general solution of the
$q$-periodic system has the following form
\begin{eqnarray}
\left(
\begin{array}{l}
\delta{\xi_j} (t) \\
\delta{\eta_j} (t)
\end{array}
\right)
&=& \Phi_j(t)
\left(
\begin{array}{l}
\delta{\xi}_j (0)\\
\delta{\eta}_j (0)
\end{array}
\right), \nonumber \\
&&\;\;\;\;\;\;\;\;j=0,1,\dots,N-1,
\label{eq:LM3}
\end{eqnarray}
Substitution of Eq.~(\ref{eq:LM3}) into Eq.~(\ref{eq:LM2}) leads to an
initial-value problem to determine $\Phi_j(t)$,
\begin{equation}
{\dot \Phi}_j(t) = J_j(t) \Phi_j(t),~\Phi_j(0)=I.
\label{eq:MCWEQ}
\end{equation}
Each $2 \times 2$ matrix $M_j$ $[\equiv \Phi_j(q)]$, which is
obtained through integration of Eq.~(\ref{eq:MCWEQ}) over the period
$q$, determines the stability of the q-periodic synchronous orbit
against the $j$th-mode perturbation.

The characteristic equation of each matrix $M_j$
$(j=0,1,\dots,N-1)$ is
\begin{equation}
\lambda_j^2 - {\rm tr}\, M_j \, \lambda_j + {\rm det} \, M_j =0,
\end{equation}
where ${\rm tr} M_j$ and ${\rm det} M_j$ denote the trace and
determinant of $M_j$, respectively.
The eigenvalues, $\lambda_{j,1}$ and $\lambda_{j,2}$, of $M_j$ are
called the Floquet (stability) multipliers, which
characterize the stability of the synchronous $q$-periodic orbit
against the $j$th-mode perturbation. Since the $j=0$ case corresponds
to the synchronous mode, the first pair
of Floquet multipliers $(\lambda_{0,1},\lambda_{0,2})$ is called the
pair of synchronous Floquet multipliers. On the other hand, all the
other pairs of Floquet multipliers are called the pairs of
asynchronous Floquet multipliers, because all the other cases of
$j \neq 0$ correspond to asynchronous modes.

As shown in \cite{Lef2}, ${\rm det}\,M_j$ is given by
\begin{equation}
{\rm det}\,M_j = e^{\int_0^q {\rm tr}\,J_j dt}=
e^{-2 \pi \beta \Omega q}.
\label{eq:MCDet}
\end{equation}
Note that all the matrices $M_j$'s have the same constant Jacobian
determinant (less than unity). Accordingly, each pair of
Floquet multipliers $(\lambda_{j,1},\lambda_{j,2})$
$(j=0,1,\dots,N-1)$ lies either on the circle of radius
$e^{-\pi \beta \Omega q}$ or on the real axis in the complex plane.
The synchronous periodic orbit is stable against the $j$th-mode
perturbation when the pair of Floquet multipliers $(\lambda_{j,1},
\lambda_{j,2})$ lies inside the unit circle in the complex plane.
We first note that the Floquet multipliers never cross the unit
circle in the complex plane and hence Hopf bifurcations do not
occur. Consequently, the synchronous periodic orbit can lose its
stability against the $j$th mode perturbation when a Floquet
multiplier $\lambda_j$ decreases (increases) through $-1(1)$ on the
real axis.

A more convenient real quantity $R_j$, called the residue and defined
by
\begin{equation}
R_j \equiv { {1 + {\rm det} M_j - {\rm tr}M_j} \over
{2(1+{\rm det}M_j)}},\;\;j=0,1,\dots,N-1,
\label{eq:MCR}
\end{equation}
was introduced in Ref.~\cite{Kim3} to characterize stability of
periodic orbits in 2D dissipative maps with constant Jacobian
determinants. Here the first one $R_0$ is associated with the
stability against the synchronous-mode perturbation, and hence it may
be called the synchronous residue. On the other hand, all the other
ones $R_j$ $(j \neq 0)$ are called the asynchronous residues, because
they are associated with the stability against the asynchronous-mode
perturbations.

A synchronous periodic orbit is stable against the $j$th-mode
perturbation when $0 < R_j <1$ (i.e., the pair of Floquet multipliers
$(\lambda_{j,1}, \lambda_{j,2})$ lies inside the unit circle in the
complex plane). When $R_j$ decreases through $0$ (i.e., a Floquet
multiplier $\lambda_j$ increases through $1$), the synchronous
periodic orbit loses its stability via saddle-node or pitchfork
bifurcation (PFB). On the other hand, when $R_j$ increases through $1$
(i.e., a Floquet multiplier $\lambda_j$ decreases through $-1$), it
becomes unstable via period-doubling bifurcation (PDB). We also note
that a(n) synchronous (asynchronous) bifurcation takes place for $j=0$
$(j \neq 0)$. For each case of the synchronous (asynchronous) PFB and
PDB, two type of supercritical and subcritical
bifurcations occur. For the supercritical case of the synchronous
(asynchronous) PFB and PDB, the synchronous periodic orbit loses its
stability and gives rise to the birth of a pair of new stable
synchronous (asynchronous) orbits with the same period and a new
stable synchronous (asynchronous) period-doubled orbit, respectively.
However, for the subcritical case of the synchronous (asynchronous)
PFB and PDB, the synchronous periodic orbit becomes unstable by
absorbing a pair of unstable synchronous (asynchronous) orbits with
the same period and an unstable synchronous (asynchronous)
period-doubled orbit, respectively. (For more details on
bifurcations, refer to Ref.~\cite{Guckenheimer}.)

It follows from the condition (\ref{eq:CC}) that the reduced coupling
functions of Eq.~(\ref{eq:RCF}) satisfy
\begin{equation}
\sum_{l=1}^{N} G_l(x) =0.
\end{equation}
Hence the matrix (\ref{eq:JML}) for $j=0$ becomes
\begin{equation}
J_0(t)=
\left( \begin{array}{cc}
0 & \;\; 1 \\
{{\partial f(x^*,\dot{x}^*,t)} \over {\partial x^*} }
& \;\; {{\partial f(x^*,\dot{x}^*,t)} \over {\partial \dot{x}^*} }
    \end{array}
\right).
\end{equation}
This is just the linearized Jacobian matrix for the case of the
uncoupled inverted pendulum \cite{Kim1}. Hence the synchronous residue
$R_0$ becomes the same as the residue of the uncoupled inverted
pendulum, i.e., it depends only on the amplitude $A$. While there is
no coupling effect on $R_0$, the coupling affects all the other
asynchronous residues $R_j$ $(j \neq 0)$.

In case of the global coupling of Eq.~(\ref{eq:GC}),
the reduced coupling functions become:
\begin{equation}
{G_l}(x) = \left \{
 \begin{array}{l}
  (1-N) G(x)\;\;\;\; {\rm for}\;l=1, \\
  \;\;\;\;\;\;G(x)\;\;\;\;\;\;\;\;\;\;{\rm for}\;l \neq 1,
  \end{array}
  \right.
\end{equation}
where $G(x)= {{\frac{c}{N}}} u'(x)$.
Substituting $G_l$'s into the second term of the $(2,1)$ entry of the
matrix $J_j(t)$ of Eq.~(\ref{eq:JML}), we have:
\begin{equation}
{\sum_{l=1}^{N}} G_l(x) e^{2 \pi i(l-1)j/N} =
\left \{ \begin{array}{l}
          \;\;\;\;\;0\;\;\;\;\;\;\;{\rm for}\;\; j=0, \\
          -c\, u'(x)\;\;{\rm for}\;\; j \neq 0.
         \end{array}
\right.
\label{eq:SE2}
\end{equation}
Hence all the asynchronous residues $R_j$ $(j \neq 0)$ become the
same, i.e., $R_1  =  \cdots = R_{N-1}$.
Consequently, there exist only two independent residues $R_0$ and
$R_1$, independently of $N$.

We next  consider the non-global  coupling of  the form (\ref{eq:GCF})
and define
\begin{equation}
G(x) \equiv {\frac {c} {2K+1}} u'(x),
\end{equation}
where $1 \leq K \leq {{{\frac {N-2} {2}}}}\;
({{\frac {N-3} {2}}})$ for even (odd) $N$ larger than 3.
Then we have
\begin{equation}
{G_l}(x) = \left \{
 \begin{array}{l}
  -2 K G(x)\;\;\; {\rm for}\;l=1, \\
  \;\;\;\;\;G(x)\;\;\;\;\;\; {\rm for}\;  2 \leq l \leq  1+K \;\; {\rm
or} \\
  \;\;\;\;\;\;\;\;\;\;\;\;\;\;\;\;\;\;\;{\rm  for}\;N+1-K \leq  l \leq
N, \\
  \;\;\;\;\;\;\;0\;\;\;\;\;\;\;\;\;\;{\rm otherwise.}
  \end{array}
  \right.
\end{equation}
Substituting the reduced coupling functions into
the  matrix $J_j(t)$,  the  second term  of the
$(2,1)$ entry of $J_j(t)$ becomes:
\begin{equation}
{\sum_{l=1}^{N}} G_l(x) e^{2 \pi i(l-1)j/N}
= - {S_N}(K,j) c\, u'(x),
\label{eq:SE1}
\end{equation}
where
\begin{equation}
{S_N}(K,j) \equiv {4 \over {2K+1}} {\sum_{k=1}^{K}}
\sin^2 {{\pi jk} \over {N}}
= 1- {\frac {\sin(2K+1) {{\frac{\pi j}{N}}}}
{(2K+1) \sin{{\frac{\pi j}{N}}}}}.
\label{eq:SF}
\end{equation}
Hence, unlike the global-coupling case, all the asynchronous residues
vary depending on the coupling range $K$ as well as on  the mode
number  $j$.
Since $S_N(K,j)  =  S_N(K,N-j)$, the residues satisfy
\begin{equation}
R_j = R_{N-j},\;\;j=0,1,\dots,N-1.
\end{equation}
Thus it is sufficient to consider only the case of
$0 \leq j \leq {N \over 2}$ $({{N-1} \over 2})$ for even (odd) $N$.
Comparing the expression in Eq.~(\ref{eq:SE1}) with that in
Eq.~(\ref{eq:SE2}) for $j \neq 0$, one can easily see that
they are the same except for the factor $S_N (K,j)$. Consequently,
making a change of the coupling parameter
${c \rightarrow {c \over {S_N (K,j)}}}$, the residue $R_j$ for the
non-global coupling case  of range $K$  becomes the same as that
for the global-coupling case.

When the synchronous residue $R_0$ of a synchronous periodic orbit
increases through $1$, the synchronous periodic orbit loses its
stability via synchronous supercritical PDB, giving rise to the
birth of a new synchronous period-doubled orbit.
Here we are interested in such synchronous supercritical PDB's.
Thus, for each mode with nonzero index $j$ we consider a region in
the $A-c$ plane, in which the synchronous periodic orbit is stable
against the perturbations of both modes with indices $0$ and $j$.
This stable region is bounded by four bifurcation curves determined
by the equations $R_0 = 0,\,1$ and $R_j=0,\,1$, and it will be
denoted by $U_N$.

For the case of global coupling, those stable regions coincide,
irrespectively of $N$ and $j$, because all the asynchronous residues
$R_j$'s $(j \neq 0)$ are the same, independently of $N$. The stable
region for this global-coupling case will be denoted by $U_G$. Note
that $U_G$ itself is just the stability region of the synchronous
periodic orbit, irrespectively of $N$, because the synchronous
periodic orbit is stable against the perturbations of all synchronous
and asynchronous modes in the region $U_G$. Thus the stability diagram
of synchronous orbits of period $2^n$ $(n=0,1,2,\dots)$ in the $A-c$
plane becomes the same, independently of $N$.

However, the stable region $U_N$ varies depending on the coupling
range $K$ and the mode number $j$ for the nonglobal-coupling cases,
i.e., $U_N=U_N(K,j)$. To find the stability region of a synchronous
periodic orbit in the $N$ coupled inverted pendulums with a given $K$,
one may start with the stability region $U_G$ for the global-coupling
case. Rescaling the coupling parameter $c$ by a scaling factor
$1 \over S_N(K,j)$ for each nonzero $j$, the stable region $U_G$ is
transformed into a stable region $U_N(K,j)$. Then the stability
region of the synchronous periodic orbit is given by the intersection
of all such stable regions $U_N$'s.

Finally, we briefly discuss Lyapunov exponents of a synchronous orbit
in the Poincar\'{e} map $P$, characterizing the mean exponential
rate of divergence of nearby orbits \cite{Lyapunov}. As shown in
Eq.~(\ref{eq:LM1}), all the synchronous and asychronous modes of a
perturbation to a synchronous orbit becomes decoupled. Hence, each
matrix $M_j$ $[\equiv \Phi_j(1)]$ with $q=1$ determines the pair of
Lyapunov exponents $(\sigma_{j,1},\sigma_{j,2})$ $(j=0,1,...,N-1)$,
characterizing the average exponential rates of divergence of
the $j$th mode perturbation, where $\sigma_{j,1} \geq \sigma_{j,2}.$
Since each $M_j$ has the same constant Jacobian determinant
(i.e., ${\rm det} M_j = e^{- 2 \pi \beta \Omega}$), each pair of
Lyapunov exponents satisfies $\sigma_{j,1}+\sigma_{j,2}=- 2 \pi
\beta \Omega$. Note also that the first pair of synchronous Lyapunov
exponents $(\sigma_{0,1},\sigma_{0,2})$ is just the pair of the
Lyapunov exponents of the uncoupled inverted pendulum \cite{Kim1},
and the coupling affects only all the other pairs of asynchronous
Lyapunov exponents $(\sigma_{j,1},\sigma_{j,2})$ $(j \neq 0)$.
Furthermore, all the pairs of the asynchronous Lyapunov exponents
for the global-coupling case become the same [i.e., $(\sigma_{1,1},
\sigma_{1,2}) = \cdots = (\sigma_{N-1,1},\sigma_{N-1,2})]$, as in
the case of the asynchronous residues.

\section{Critical Scaling Behaviors of Period Doublings}
\label{sec:CSB}

In this section, by varying the two parameters $A$ and $c$, we study
the critical scaling behaviors of synchronous PDB's in the $N$
symmetrically coupled inverted pendulums for $\beta=0.2$ and
$\Omega=0.1$. It is found that the critical behaviors depend on the
coupling range. In the global-coupling case, in which each inverted
pendulum is coupled to all the other ones with equal coupling
strength, the zero-coupling critical point and an infinity of critical
line segments constitute the same critical set, independently of $N$.
However, for any other nonglobal-coupling cases, the structure of the
critical set becomes different from that for the global-coupling case,
because of a significant change in the stability diagram of the
synchronous $2^n$-periodic orbits $(n=0,1,2, \dots)$.
The critical scaling behaviors on the critical set are found to be the
same as those for the abstract system of the coupled 1D maps
\cite{Kim2}. We thus consider separately two kinds of couplings, the
global- and nonglobal-coupling cases.

\subsection{Global Coupling}
\label{subsec:GC}

We first study the $N$ globally-coupled inverted pendulums with the
coupling function of the form (\ref{eq:GC}). As shown in
Sec.~\ref{sec:SBL}, a synchronous periodic orbit is stable when all
its residues $R_j$ $(j=0,1,...,N-1)$ defined in Eq.~(\ref{eq:MCR}) 
lie between 0 and 1 (i.e., $0< R_j <1$). Here $R_0$ is the synchronous 
residue determining the stability against the synchronous-mode 
perturbation, while all the other ones $R_j$ $(j \neq 0)$ are the 
asynchronous residues determining the stability against the 
asynchronous-mode perturbations. For the globally-coupled case,  
all the asynchronous residues become the same, independently of $j$,
and hence only one independent asynchronous residue (e.g., $R_1$) 
exists. Accordingly, the stability region of a synchronous periodic
orbit becomes bounded by four bifurcation lines determined by the
equations $R_0=0,\,1$ and $R_1=0,\,1$. Here the $R_0=0$ 
and $1$ ($R_1=0$ and $1$) lines correspond to the synchronous 
(asynchronous) PFB and PDB lines, respectively. In such a way, we
obtain the stability diagram of the synchronous $2^n$-periodic orbits 
$(n=0,1,2, \dots)$ in the $A-c$ plane. Note also that the stability 
diagram becomes the same, independently of $N$, because all the 
asynchronous residues $R_j$ $(j \neq 0)$ for each synchronous orbit 
are also the same, irrespectively of $N$. Consequently, the structure 
of the critical set and the critical behaviors for the global-coupling
case become the same, independently of $N$.

As an example, we consider a linearly coupled case in which the
coupling function (\ref{eq:GC}) is
\begin{equation}
g(x_1,\dots,x_N)  =
  c\,[ {\frac  {1}{N}} {\sum_{m=1}^{N}} x_m  - x_1].
 \label{eq:LC}
\end {equation}
As in the uncoupled inverted pendulum \cite{Kim1}, the coupled
inverted pendulums exhibit multiple period-doubling transitions to
chaos. Here we study the first three period-doubling transitions to
chaos. For each period-doubling transition to chaos, the zero-coupling
critical point and an infinity of critical line segments constitute
the critical set in the $A-c$ plane. Three kinds of critical behaviors
associated with the scaling of the coupling parameter $c$ are found on
the critical set, while the critical scaling behavior of the amplitude
$A$ is always the same as that of the uncoupled inverted pendulum.
Note that the structure of the critical set and the critical behaviors
for the coupled inverted pendulums are found to be the same as those 
for the coupled 1D maps \cite{Kim2}.

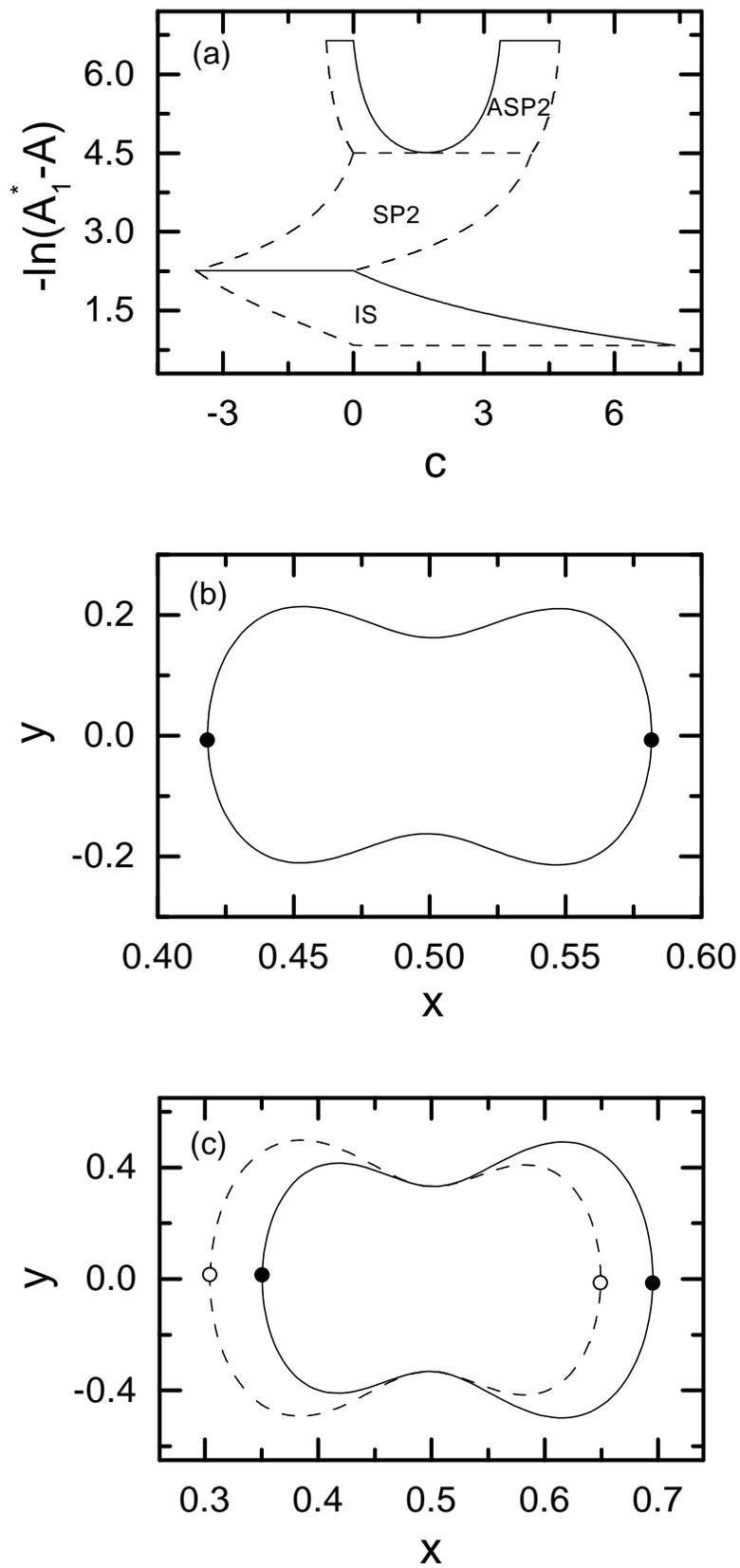
\begin{figure}
\caption{(a) Stability diagram of the synchronous orbits of low period
          $q=1,2$ in $N$ linearly-coupled inverted pendulums with the
          global coupling. Here $A^*_1$ $(=0.575\,154\, \cdots)$ is 
          just the first period-doubling transition point of the 
          uncoupled inverted pendulum. The stable regions of the 
          inverted stationary point, a symmetric 2-periodic orbit, and 
          an asymmetric 2-periodic orbit are denoted by the SP, the 
          SP2, and the ASP2, respectively. The horizontal 
          (non-horizontal) solid and dashed boundary lines correspond 
          to synchronous (asynchronous) PDB and PFB lines, 
          respectively. (b) Phase portraits for $A=0.5$. The phase flow 
          of a symmetric 2-periodic orbit born via synchronous 
          supercritical PDB is denoted by a solid curve, and its 
          Poincar\'{e} maps are represented by the solid 
          circles. (c) Phase portraits for $A=0.57$. The phase flows
          of a conjugate pair of asymmetric 2-periodic orbits born
          via synchronous supercritical PFB are shown: one is denoted
          by a solid curve, while the other one is denoted by a dashed
          curve. Their Poincar\'{e} maps are also represented by the
          solid and open circles, respectively.
     }
\label{SD1}
\end{figure}

Figure \ref{SD1}(a) shows the stability diagram of the synchronous 
orbits with low period $q=1,2$. The stable region of a synchronous 
orbit is bounded by its PDB and PFB lines. The horizontal 
(non-horizontal) solid and dashed boundary lines correspond to 
synchronous (asynchronous) PDB and PFB lines, respectively. Each 
bifurcation may be supercritical or subcritical.

We first consider the bifurcations associated with stability of the
synchronous inverted stationary point, corresponding to the
vertically-up configuration (i.e., $x_1(t)= \cdots = x_N (t) \equiv
x^*(t) = {1 \over 2}$ and $y_1(t)= \cdots = y_N(t) \equiv y^*(t)=0$).
The inverted state is a symmetric one with respect to the inversion
symmetry $S$. Its stability region is denoted by the IS in
Fig.~\ref{SD1}(a). For the unforced case of $A=0$, the inverted state 
is obviously unstable. However, when crossing the horizontal dashed
boundary line of the IS, its first resurrection occurs, i.e., it
becomes stabilized with birth of a pair of unstable synchronous
asymmetric orbits with period $1$ via subcritical PFB. (For more
details on the resurrection of the inverted state, refer to
Ref.~\cite{Kim1}.) This stabilized inverted state destabilizes again
through asynchronous PDB and PFB when the nonhorizontal solid and
dashed boundary curves are crossed, respectively. However, it
becomes unstable via synchronous supercritical PDB when crossing the
horizontal solid boundary line, and gives rise to the birth of a new
synchronous orbit of period $2$. This new synchronous $2$-periodic
orbit also is a symmetric one with respect to the inversion symmetry
$S$, as shown in Fig.~\ref{SD1}(b) and its stable region is denoted 
by the SP2 in Fig.~\ref{SD1}(a). This synchronous symmetric orbit of 
period $2$ loses its stability through asynchronous PFB's when 
crossing the non-horizontal dashed boundary curves. However, it 
becomes unstable via synchronous supercritical PFB when the 
horizonatl dashed boundary line is crossed, and consequently a pair 
of new stable synchronous orbits with the same period $2$ appears. 
Note that the new pair of synchronous orbits is a conjugate pair of 
asymmteric orbits with respect to the inversion symmetry $S$, which
is shown in Fig.~\ref{SD1}(c). That is, the inversion symmetry is 
broken due to the symmetry-breaking PFB. The stable region of the 
asymmetirc orbits of period $2$ is denoted by the ASP2 in 
Fig.~\ref{SD1}(a). Each synchronous asymmetric $2$-periodic
orbit becomes unstable via synchronous supercritical PDB when the
horizontal solid boundary line is crossed, and gives rise to the
birth of a new synchronous asymmetric $4$-periodic orbit. Here we
are interested in such synchronous supercritical PDB's.

Figure \ref{SD2} shows the stability diagram of synchronous
asymmetric orbits born by synchronous supercritical PDB's.
Each synchronous asymmetric orbit of level $n$ (period $2^n$,
$n=1,2,3,\dots$) loses its stability at the horizontal solid boundary
line of its stable region via synchronous supercritical PDB, and gives
rise to the birth of a synchronous asymmetric period-doubled orbit of
level $n+1$. Such an infinite sequence ends at a finite value of
$A^*_1 = 0.575\,154\, \cdots$, which is just the first period-doubling
transition point of the uncoupled inverted pendulum \cite{Kim1}.
Consequently, a synchronous quasiperiodic orbit, whose maximum
synchronous Lyapunov exponent is zero (i.e., $\sigma_{0,1}=0$), exists
on the $A=A^*_1$ line.

We examine the treelike structure of the stability diagram in
Fig.~\ref{SD2}, which consists of an infinite pile of $U$-shape
regions and rectangular-shape regions. Note that the treelike
structure is asymptotically the same as that in the coupled 1D maps
\cite{Kim2}. The $U$-shape branching is repeated at one side of each
$U$-shape region, including the $c=0$ line segment. The branching side
will be referred to as the zero $c$ side. However, the other side of
each $U$-shape region grows like a chimney without any further
branchings (as an example, see the branch in Fig.~\ref{SD2}(b) starting
from the right side of the $U$-shape region of the ASP2).
As in the coupled 1D maps \cite{Kim2}, this rule governs the
asymptotic behavior of the treelike structure.

A sequence of connected stability regions with increasing period is
called a ``period-doubling route'' \cite{Kim2}. There are two kinds of
period-doubling routes. The sequence of the $U$-shape regions with the
zero $c$ sides converges to the zero-coupling point $c=0$ on the
$A=A^*_1$ line. It will be referred to as the $U$ route. On the other
hand, a sequence of rectangular regions in each chimney converges to a
critical line segment on the $A=A^*_1$ line. For example, the sequence
of the rectangular regions in Fig.~\ref{SD2}(b) converges to a
critical line segment joining the left end point $c_l$
$(=3.427\,742 \cdots)$ and the right end point $c_r$
$(=4.796\,277 \cdots)$ on the $A=A^*_1$ line. This kind of route will
be called a $C$ route. Note that there are infinitely many $C$ routes,
while the $U$ route converging to the zero-coupling critical point
$(A^*_1,0)$ is unique. Hence, an infinite number of critical line
segments, together with the zero-coupling critical point, constitute
the critical set.

We now study the critical behaviors on the critical set.
First, consider the case of the $U$ route ending at the zero-coupling
critical point. We follow the synchronous orbits of period $q=2^n$ up
to level $n=8$ in the $U$ route, and obtain a self-similar sequence of
parameters $(A_n,c_n)$, at which each orbit of level $n$ has some
given synchronous and asynchronous residues $R_0$ and $R_1$
$(=R_2 = \cdots = R_{N-1})$ (e.g., $R_0=1$ and $R_1=0$).
Then the sequence $\{ (A_n,c_n) \}$ converges
geometrically to the zero-coupling critical point $(A^*_1,0)$. As in
the uncoupled inverted pendulum \cite{Kim1}, the sequence $\{ A_n \}$
obeys a scaling law,
\begin{equation}
 \Delta A_n \sim \delta^{-n}\;\; {\rm for \; large\;} n,
 \label{eq:ASL}
 \end{equation}
where $\Delta A_n = A_n - A_{n-1}$ and $\delta \simeq 4.67$.
The value of the scaling factor $\delta$ agrees well with the
Feigenbaum constant $(=4.669 \cdots)$ of the 1D
map \cite{Feigenbaum}. We also note that the sequence $\{ c_n \}$
obeys a scaling law,
\begin{equation}
 \Delta c_n \sim \mu^{-n}\;\; {\rm for \; large\;} n,
 \label{eq:cSL}
 \end{equation}
where $\Delta c_n = c_n - c_{n-1}$
The sequence of the scaling factor $\{ \mu_n \}$
$(=\Delta c_n / \Delta c_{n+1})$ of level $n$ is listed in the second
column of Table \ref{table1} and converges to a constant $\mu$
$(\simeq -2.5)$, which agrees well with the coupling-parameter scaling
factor $\alpha$  $(=-2.502\cdots)$ of the coupled 1D maps
near the zero-coupling critical point \cite{Kim2}. It has been also
shown in \cite{Kim2} that the scaling factor $\alpha$ is just the
largest relevant ``coupling eigenvalue'' (CE) of the zero-coupling
fixed map of the renormalization transformation for the case of the
coupled 1D maps.

We also study the coupling effect on the asynchronous residue
$R_{1,n}$ of the synchronous orbit of period $2^n$ near the
zero-coupling critical point $(A^*_1 , 0)$. Figure \ref{AR1}
shows three plots of $R_{1,n} (A^*_1,c)$ versus $c$ for
$n=5,6$ and $7$. For $c=0$, $R_{1,n}$ converges to a constant
$R_1^*$ $(=1.300\,59\dots)$, called the critical asynchronous residue,
as $n \rightarrow \infty$. However, when $c$ is nonzero $R_{1,n}$
diverges as $n \rightarrow \infty$, i.e., its slope $S_n$
$\displaystyle{ \left.
(\equiv  {{\partial R_{1,n}} \over {\partial c}}
\right|_{(A^*_1 ,0)})
}$ at the zero-coupling critical point diverges as
$n \rightarrow \infty$.

As in the scaling for the coupling parameter $c$, the sequence
$\{ S_n \}$ also obeys a scaling law,
\begin{equation}
S_n \sim \nu^n \;\;\;{\rm for\;large\;}n.
\label{eq:SLSL}
\end{equation}
The scaling factor $\nu_n$ $(= S_{n+1} / S_n)$ of level $n$ is listed
in the third column of Table \ref{table1} and converges to a constant
$\nu$ $(\simeq -2.5)$ as $n \rightarrow \infty$. Note also that the
value of $\nu$ agrees well with that of the largest relevant CE
$\alpha$ of the zero-coupling fixed map.

We next consider the cases of $C$ routes, each of which converges to a
critical line segment. Two kinds of additional critical behaviors are
found at each critical line segment; the one critical behavior exists
at both ends and the other critical behavior exists at interior points.
In each $C$ route, there are two kinds of self-similar sequences of
parameters $(A_n,c_n)$, at which each synchronous orbit of level $n$
has some given synchronous and asynchronous residues $R_0$ and $R_1$;
the one converges to the left end point of the critical line segment
and the other converges to the right end point. As an example,
consider the $C$ route in Fig.~\ref{SD2}(b), which converges to the
critical line segment with two ends $(A^*_1,c_l)$ and $(A^*_1,c_r)$.
We follow, in the $C$ route, two self-similar sequences of
parameters, one converging to the left end and the other converging
to the right end. In both cases, the sequence $\{ A_n \}$ converges
geometrically to its accumulation value $A^*_1$ with the 1D scaling
factor $\delta$ $(\simeq 4.67)$ like the case of the $U$ route.
The sequences $\{ c_n \}$ for both cases also obey the scaling law,
\begin{equation}
\Delta c_n \sim \mu^{-n} \;\;{\rm for\;large}\; n,
\end{equation}
where $\Delta c_n = c_n - c_{n-1}$.
The sequence of the scaling factor $\mu_n$
$(= \Delta c_n / \Delta c_{n+1})$ of level $n$ is listed in
Table \ref{table2}, and converges to its limit value
$\mu$ $(\simeq 2)$. We also note that the value of $\mu$ agrees well
with that of the coupling-parameter scaling factor ($\nu=2$) of the
coupled 1D maps near both ends of each critical line segment
\cite{Kim2}. It has been also shown in \cite{Kim2} that the scaling
factor $\nu$ $(=2)$ is just the only relevant CE of a nonzero-coupling
fixed map of the renormalization transformation for the case of the
coupled 1D maps.

Figure \ref{AR2}(a) shows the behavior of the asynchronous residue
$R_{1,n}(A^*_1,c)$ of the synchronous orbit of period $2^n$
$(n=5,6,7)$ near the critical line segment in Fig.~\ref{SD2}(b).
Magnified views near the both ends $c_l$ and $c_r$ are also given in
Figs.~\ref{AR2}(b) and \ref{AR2}(c), respectively. For $c=c_l$ and
$c_r$, $R_{1,n}$ converges to a critical asynchronous residue $R_1^*$
$(=0)$ as $n \rightarrow \infty$, which is different from that for
the zero-coupling case. The sequence of the slope $S_n$ of $R_{1,n}$
at both ends obeys well the scaling law,
\begin{equation}
S_n \sim \nu^n \;\;{\rm for\; large\;} n.
\end{equation}
The two sequences of the scaling factors $\nu_n$ $(=S_{n+1} / S_n)$
of level $n$ at both ends are listed in Table \ref{table3}, and
converge to their limit values $\nu \simeq 2$, which agrees well with
the only CE ($\nu=2$) of the nonzero-coupling fixed map governing
the critical behavior at both ends for the case of the coupled 1D
maps. However, for any fixed value of $c$ inside the critical line
segment, $R_{1,n}$ converges to a critical asynchronous residue
$R_1^*$ $(=0.5)$ as $n \rightarrow \infty$ [see Fig.~\ref{AR2}(a)].
This case of $R_1^*=0.5$ corresponds to the superstable case of
$\lambda^*_1=0$ $(\lambda^*_1$: the critical asynchronous Floquet
multiplier) for the coupled 1D maps \cite{Kim2}, because
Eq.~(\ref{eq:MCR}) of $R$ for the case of 2D maps reduces to the
equation of $R=0.5 \times (1-\lambda)$ for the case of 1D maps. We
also note that as in the case of the coupled 1D maps, there exists
no scaling factor of the coupling parameter inside the critical line
segemnt, and hence the coupling parameter becomes an irrelevant one
at interior critical points. Thus, the critical behavior inside
the critical line segment becomes the same as that of the
uncoupled inverted pendulum (i.e., that of the 1D map),
which will be discussed in more details below. This kind of
1D-like critical behavior was found to be governed by another
nonzero-coupling fixed map with no relevant CE for the case of the
coupled 1D maps \cite{Kim2}.

There exists a synchronous quasiperiodic orbit on the $A=A^*_1$ line.
As mentioned in Sec.~\ref{sec:SBL}, its synchronous Lyapunov exponents
are the same as the Lyapunov exponents of the uncoupled inverted
pendulum, i.e., $\sigma_{0,1}=0$ and $\sigma_{0,2}=-2 \pi \beta
\Omega$. The coupling affects only the pair of asynchronous
Lyapunov exponents $(\sigma_{1,1},\sigma_{1,2})$
$[=(\sigma_{2,1},\sigma_{2,2})= \cdots = (\sigma_{N-1,1},
\sigma_{N-1,2})]$. The maximum asynchronous Lyapunov exponent
$\sigma_{1,1}$ near the critical line segment in Fig.~\ref{SD2}(b) is
shown in Fig.~\ref{ALYP1}. Inside the critical line segment
$(c_l < c < c_r)$, the synchronous quasiperiodic orbit on the
synchronization plane becomes a synchronous attractor with
$\sigma_{1,1}<0$. Since the dynamics on the synchronous attractor is
the same as that of the uncoupled inverted pendulum, the
critical maps at interior points exhibit essentially 1D-like critical
behaviors, because the critical behavior of the uncoupled inverted
pendulum is the same as that of the 1D maps \cite{Kim1}.
However, as the coupling parameter $c$ passes through $c_l$ and $c_r$,
the maximum asynchronous Lyapunov exponent $\sigma_{1,1}$ of the
synchronous quasiperiodic orbit increases from zero. Consequently,
the synchronous quasiperiodic orbit ceases to be an attractor outside
the critical line segment, and the system of the coupled inverted
pendulums is asymptotically attracted to another synchronous
rotational attractor of period $1$.

What happens beyond the first period-doubling transition point $A^*_1$
is also interesting. As in the uncoupled 1D inverted pendulum
\cite{Kim1}, with increasing the amplitute $A$ further from $A=A^*_1$,
the unstable inverted state undergoes a cascade of resurrections,
i.e., it will restabilize after it loses its stability, destabilize
again, and so forth {\it ad infinitum}. For each case of the
resurrections, an infinite sequence of PDB's leading to chaos follows.
Consequently, the coupled inverted pendulums exhibit multiple
period-doubling transitions to chaos.

As the first example, we consider the second period-doubling
transition to chaos. Figure \ref{SD3}(a) shows the second stability
diagram of the synchronous inverted stationary point and asymmetric
orbits of level $n$ (period $2^n$, $n=0,1,2,3$) in the $A-c$ plane.
When crossing the horizontal solid boundary line of its stability
region IS, the unstable inverted state restabilizes with birth of a
new unstable synchronous symmetric orbit of period $2$ via synchronous
subcritical PDB. This is the second resurrection of the inverted
state. However, when the horizontal dashed boundary line is crossed,
the stabilized inverted state becomes unstable via synchronous
supercritical PFB, which results in the birth of a conjugate pair of
synchronous asymmetric orbits with period $1$. Then each synchronous
asymmetric orbit of level $n$ becomes unstable at the horizontal solid
boundary line of its stability region via synchronous supercritical
PDB, and gives rise to the birth of a synchronous asymmetric
period-doubled orbit of level $n+1$. Such an infinite sequence
terminates at a finite value of $A^*_2 (= 3.829\,784\, \cdots)$, which
is the second period-doubling transition point of the uncoupled
inverted pendulum \cite{Kim1}. Note that the treelike structure of the
stability diagram in Fig.~\ref{SD3}(a) is essentially the same as that
in Fig.~\ref{SD2}(a). Hence, the critical set also consists of the
zero-coupling critical point and an infinite number of critical line
segments, as in the first period-doubling transition case. In
order to study the critical behaviors on the critical set, we follow
the synchronous asymmetric orbits up to level $n=7$ in the $U$ route
and in the rightmost $C$ route. It is found that the critical
behaviors are the same as those for the first period-doubling
transition case. That is, there exist three kinds of critical
behaviors at the zero-coupling critical point, both ends of each
critical line segment and interior points.

As the second example, we also consider the third period-doubling
transition to chaos. The third stability diagram of the synchronous
orbits with $q=1,2,4,8$ is shown in Fig.~\ref{SD3}(b). A synchronous
subcritical PFB occurs when crossing the horizontal dashed boundary
line of the IS. Consequently, the unstable inverted state
restabilizes with birth of a pair of unstable orbits with period $1$.
This is the third resurrection of the inverted state. However, the
stabilized inverted state becomes unstable via synchronous
supercritical PDB when the horizontal solid boundary line of the IS
is crossed, and gives rise to the birth of a symmetric $2$-periodic
orbit. The subsequent bifurcation behaviors are the same as those
for the first period-doubling transition to chaos. That is, a third
infinite sequence of synchronous supercritical PDB's follows and
ends at a finite value $A^*_3 (= 10.675\, 090\, \cdots)$, which is
the third period-doubling transition point of the uncoupled
inverted pendulum \cite{Kim1}. Note also that the treelike structure
of the third stability diagram is essentially the same as that in
Fig.~\ref{SD2}(a). Hence, the critical set is composed of the
zero-coupling critical point and an infinity of critical line
segments. Furthermore, the critical behaviors on the critical
set are found to be the same as those for the first period-doubling
transition case.

In addition to the linear-coupling case (\ref{eq:LC}), we have also
studied two other nonlinear-coupling cases,
\begin{equation}
g(x_1,\dots,x_N)  =
  c\,[ {\frac  {1}{N}} {\sum_{m=1}^{N}} x_m^n  - x_1^n], \;\;n=2,3.
\label{eq:NLC}
\end{equation}
First stability diagrams of the synchronous orbits for the cases of
the quadratic and cubic couplings are shown in Fig.~\ref{NCSD}(a)
and \ref{NCSD}(b), respectively. Their treelike structures are
essentially the same as that in Fig.~\ref{SD2}(a). Hence, the
zero-coupling critical point and an infinite number of critical line
segments constitute the critical set for each nonlinear-coupling case.
Moreover, the critical behaviors for these nonlinear-coupling cases
are also found to be the same as those for the linear-coupling case.

\subsection{Nonglobal Coupling}
\label{subsec:NGC}

 Here we study the nonglobal-coupling cases with the coupling range
 $K < {N \over 2} ({ {N-1} \over 2 })$ for even (odd) $N$.
 The structure of the critical set becomes different from that for
 the  global-coupling case, because of a significant change in the
 stability diagram of the synchronous orbits with period $2^n$
 $(n=0,1,2, \dots)$, as will be seen below.

 As an example, we consider a linearly-coupled, nearest-neighbor
 coupling case with $K=1$, in which the coupling function is
\begin{equation}
g(x_1,\dots, x_N) = {c \over 3} (x_2 + x_N - 2 x_1) \,\,{\rm for}
\, N>3.
\end{equation}
 As shown in Sec.~\ref{sec:SBL}, the stable region $U_N$, in which a
 synchronous orbit is stable against the perturbations of both
 modes with indices $0$ and $j$ $(\neq 0)$, varies depending on the
 mode number $j$, because the asynchronous residue $R_j$ $(j \neq 0)$
 depends on $j$. To find the stability region of the synchronous
 orbit, one can start with the stability region $U_G$ for the
 global-coupling case. Rescaling the
 coupling parameter $c$ by a scaling factor $1/S_N(1,j)$
 [$S_N(K,j)$ is given in Eq.~(\ref{eq:SF})], the stable region $U_G$
 is transformed into a stable region $U_N(1,j)$. Then the stability
 region of the synchronous orbit is given by the intersection of all
 such stable regions $U_N$'s.

As an example, we consider the case with $N=4$. Figure \ref{MCSD}
shows the stability regions of the synchronous asymmetric
$2^n$-periodic $(n=1,2,3,4)$ orbits. Note that the scaling factor
$1 \over {S_4(1,j)}$ has its minimum value $3 \over 4$ at $j=2$.
However, for each synchronous orbit, $U_4(1,2)$ itself cannot be the
stability region, because bifurcation curves of different
modes with nonzero indices intersect one another. We now examine
the structure of the stability diagram in Fig.~\ref{MCSD}, starting
from the left side of the stability region of the synchronous
asymmetric orbit of level $1$ $(n=1)$. For the case of level
$2$ $(n=2)$, the zero $c$ side of $U_4(1,2)$ including a $c=0$ line
segment remains unchanged, whereas the other side becomes flattened
by the bifurcation curve of the asynchronous mode with $j=1$
\cite{Rem}. Due to the successive flattening with increasing level
$n$, a significant change in the stability diagram occurs. Of the
infinite number of period-doubling routes for the global-coupling
case, only the $U$ route ending at the zero-coupling critical point
remains. Thus only the zero-coupling point is left as a critical
point in the parameter plane.

Consider a self-similar sequence of parameters $(A_n,c_n)$, at which
the synchronous orbit of period $2^n$ has some given residues, in
the $U$ route for the global-coupling case. Rescaling the coupling
parameter with the minimum scaling factor $1 \over S_4(1,2)$
$(=0.75)$, the sequence is transformed into a self-similar one for
the $N=4$ case of nearest-neighbor coupling. Hence, the critical
behavior near the zero-coupling critical point becomes the same as
that for the global-coupling case.

The results for the nearest-neighbor coupling case with $K=1$ extends
to all the other nonglobal-coupling cases with
$1 < K < {N \over 2} ({{N-1} \over 2})$
for even (odd) $N$. For each nonglobal-coupling case with $K>1$, we
first consider a mode with index $j_{\rm min}$ for which the scaling
factor $1 \over S_N(K,j)$ becomes the smallest one and the stability
region $U_N(K,j_{\rm min})$ including a $c=0$ line segment. Here the
value of $j_{\rm min}$ varies depending on the range $K$. Like the
$K=1$ case, the zero $c$ side of $U_N(K,j_{\rm min})$ including the
$c=0$ line segemnt remains unchanged, whereas the other side becomes
flattened by the bifurcation curves of the other modes with nonzero
indices. Thus the overall shape of the stability diagram of the
$2^n$-periodic $(n=1,2,3, \cdots)$ orbits born via synchronous
supercritical PDB's becomes essentially the same as that for the
nearest-neighbor coupling case. Consequently, only the $U$
route ending at the zero-coupling critical point is left as a
period-doubling route, and the critical behavior near the
zero-coupling critical point is also the same as that for the
global-coupling case.

\section{Summary}
\label{sec:Sum}

The critical behaviors of period doublings in the system of $N$
symmetrically coupled inverted pendulums have been investigated by
varying the two parameters $A$ and $c$. As in the single inverted
pendulum \cite{Kim1}, the coupled inverted pendulums exhibit multiple
period-doubling transitions to chaos with increasing $A$. We have
studied the first three period-doubling transitions to chaos. For
each period-doubling transition to chaos, it has been found that the
critical behaviors vary depending on whether or not the coupling
is global. For the global-coupling case the zero-coupling critical
point and an infinity of critical line segments constitute the same
critical set in the $A-c$ plane, independently of $N$. However, for
any other nonglobal-coupling cases the structure of the critical set
becomes different from that for the global-coupling case, because of a
significant change in the stability diagram of $2^n$-periodic orbits
$(n=0,1,2,  \dots)$. The critical scaling behaviors on the critical set
have been also found to be the same as those for the abstract
system of the coupled 1D maps \cite{Kim2}.

\acknowledgments
Some part of this work has been done while S.Y.K. visited the Centre
for Nonlinear Studies of the Hong Kong Baptist University. This work
was supported by the Basic Science Research Institute Program,
Ministry of Education, Korea, Project No. BSRI-97-2401 (S.Y.K.) and
in part by grants from the Hong Kong Research Grants Council (RGC) and
the Hong Kong Baptist University Faculty Research Grant (FRG).

\begin{table}
\caption{For the case of the $U$ route, the scaling factors
$\mu_n$ and $\nu_n$ in the scaling for the coupling parameter
and the slope of the asynchronous residue at the
zero-coupling critical point are shown in the second and third
columns, respectively.}
\label{table1}
\begin{tabular}{ccc}
$n$ & $\mu_n$ & $\nu_n$  \\
\tableline
4 & -3.517 & -2.958 \\
5 & -2.904 & -2.627 \\
6 & -2.530 & -2.480 \\
7 & -2.495 & -2.522
\end{tabular}
\end{table}

\begin{table}
\caption{We followed, in the $C$ route in Fig.~2(b), two
         self-similar sequences of parameters $(A_n,c_n)$, at which
         the pair of residues $(R_{0,n}, R_{1,n})$ of the synchronous
         orbit with period $2^n$ is $(1,0.1)$. They converge to both
         ends of the critical line segment. The scaling factors of
         the coupling paramter at the left and right ends are shown
         in the second and third columns, respectively. In both
         cases the scaling factors seem to converge to the same limit
         value $\mu \simeq 2$.
        }
\label{table2}
\begin{tabular}{ccc}
$n$ & $\mu_n$ & $\mu_n$  \\
\tableline
4 & 3.66 & 3.93 \\
5 & 2.81 & 3.04 \\
6 & 2.02 & 2.15 \\
7 & 1.93 & 1.99
\end{tabular}
\end{table}

\begin{table}
\caption{The scaling factors $\nu_n$'s in the scaling for the
slope of the asynchronous residue at the left and right ends of the
critical line segment in Fig.~2(b) are shown in the second and
third columns, respectively.
        }
\label{table3}
\begin{tabular}{ccc}
$n$ & $\nu_n$ & $\nu_n$  \\
\tableline
4 & 2.528 & 2.525 \\
5 & 2.071 & 2.072 \\
6 & 2.001 & 2.001 \\
7 & 2.000 & 2.000
\end{tabular}
\end{table}

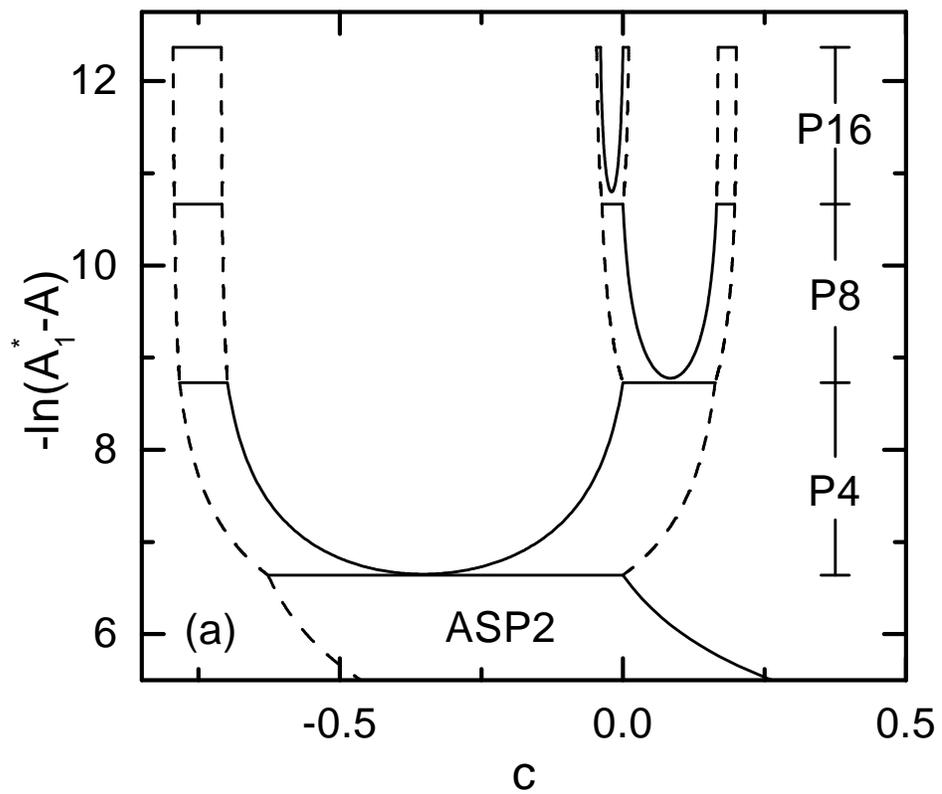
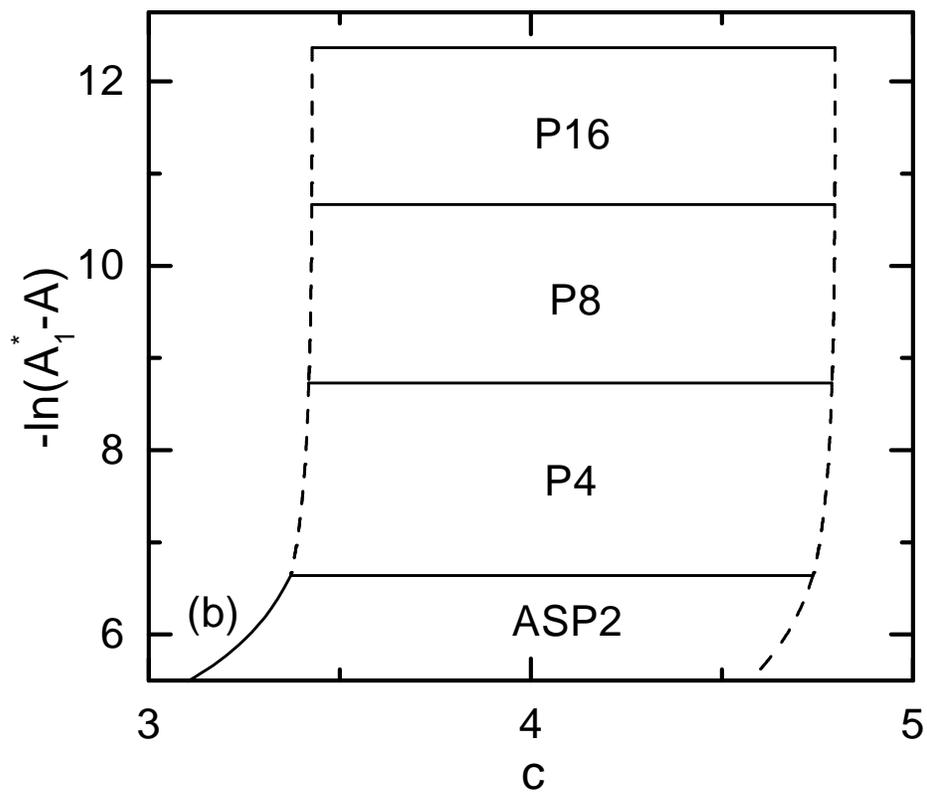
\begin{figure}
\caption{Stability diagram of synchronous asymmetric
         $2^n$-periodic ($n=1,2,3,4$) orbits of level $n$ born via
         synchronous supercritical PDB's. PN denotes the stable region
         of an asymmetric orbit of period N (N$=2,4,8,16$).
         The solid and dashed boundary lines represent the same
         as those in Fig.~\ref{SD1}. The stability diagram starting
         from the left (right) side of the ASP2 is shown in (a) [(b)].
         Note its treelike structure. 
     }
\label{SD2}
\end{figure}

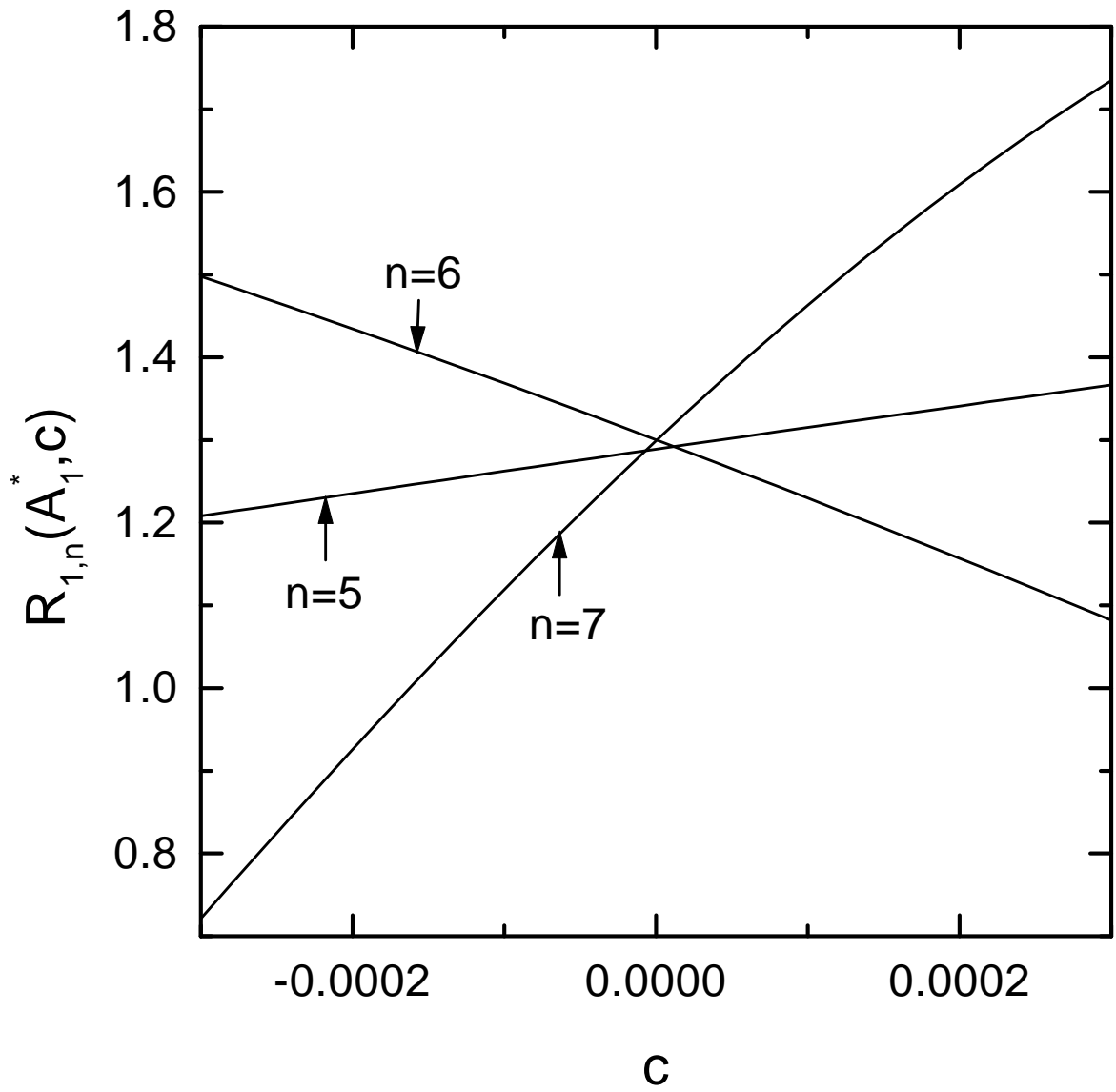
\begin{figure}
\caption{Plots of the asynchronous residue $R_{1,n}(A^*_1,c)$ versus
     $c$ near the zero-coupling critical point for $n=5,6,7$.
     }
\label{AR1}
\end{figure}

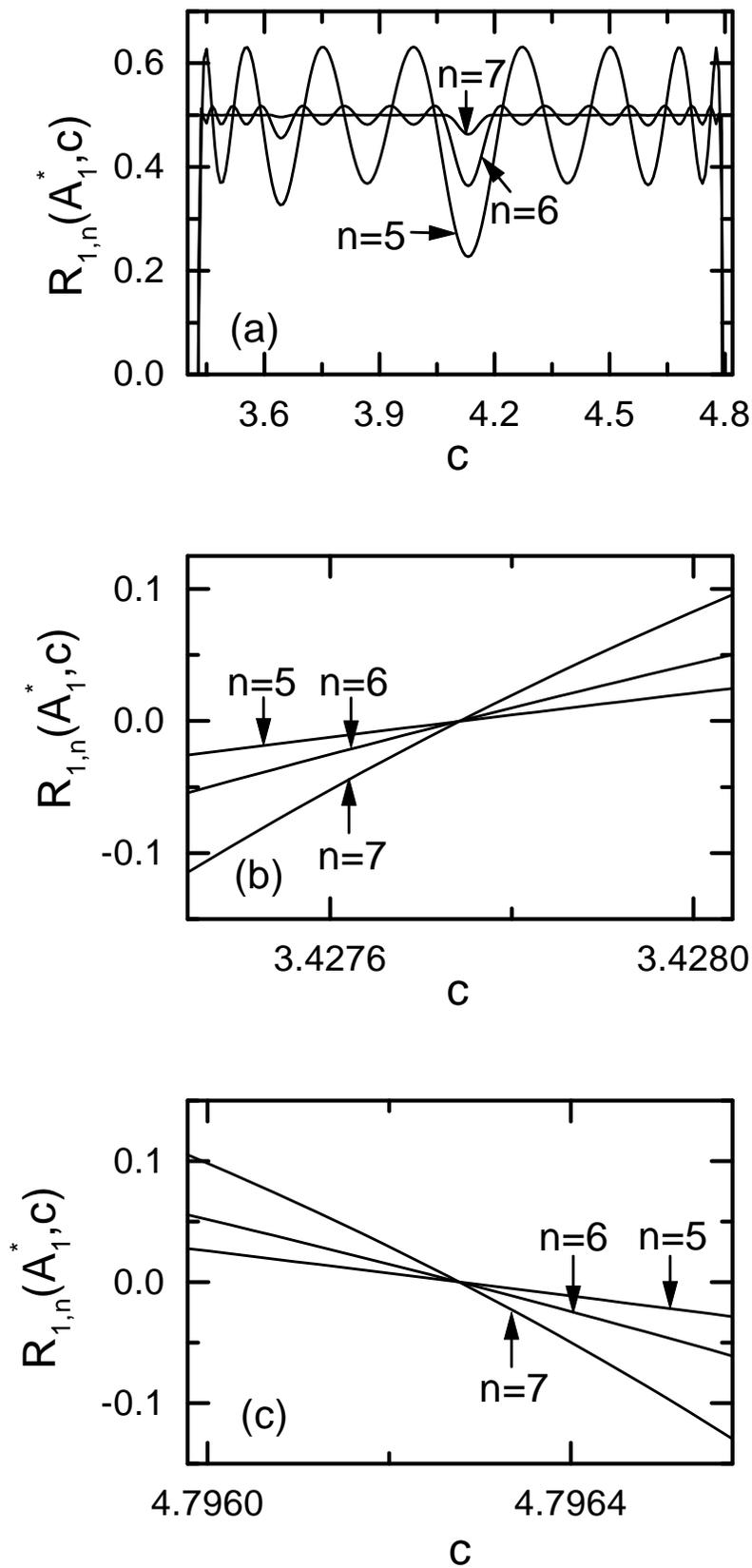
\begin{figure}
\caption{(a) Plots of the asynchronous residue $R_{1,n}(A^*_1,c)$
          versus $c$ near the critical line in Fig.~2(b) for
          $n=5,6,7$. Their magnified views near the both ends $c_l$
          and $c_r$ are also given in (b) and (c), respectively.
     }
\label{AR2}
\end{figure}

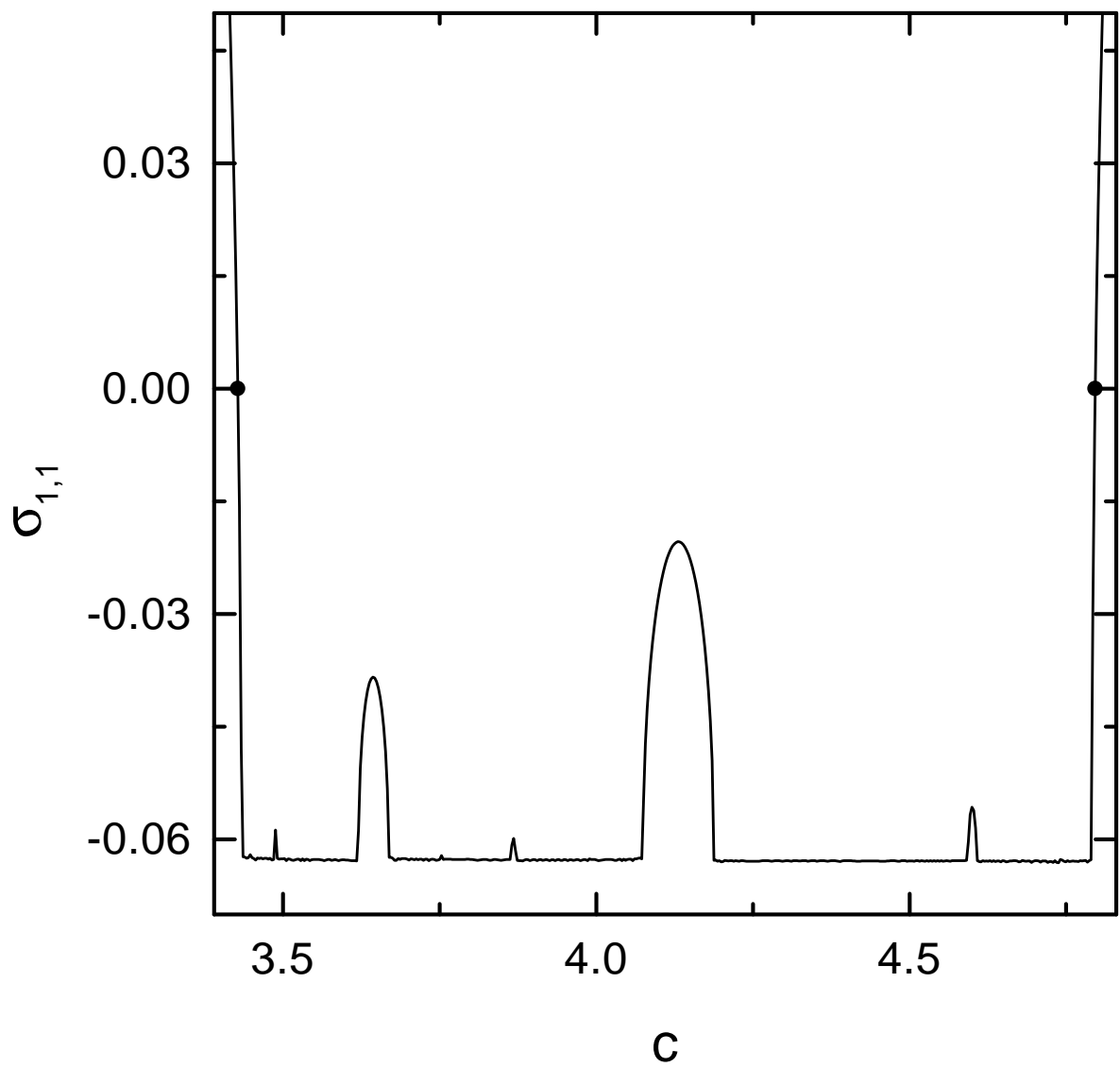
\begin{figure}
\caption{Plot of the maximum asynchronous Lyapunov exponent
         $\sigma_{1,1}$ of the synchronous quasiperiodic orbit near
         the critical line in Fig.~2(b). This plot consists
         of $450$ $c$ values, each of which is obtained by iterating
         the Poincar{\'e} map $P$ $20\,000$ times to eliminate
         transients and then averaging over another $5000$ iterations.
         The values of $\sigma_{1,1}$ at both ends of the critical
         line are zero, which are denoted by solid circles.
     }
\label{ALYP1}
\end{figure}

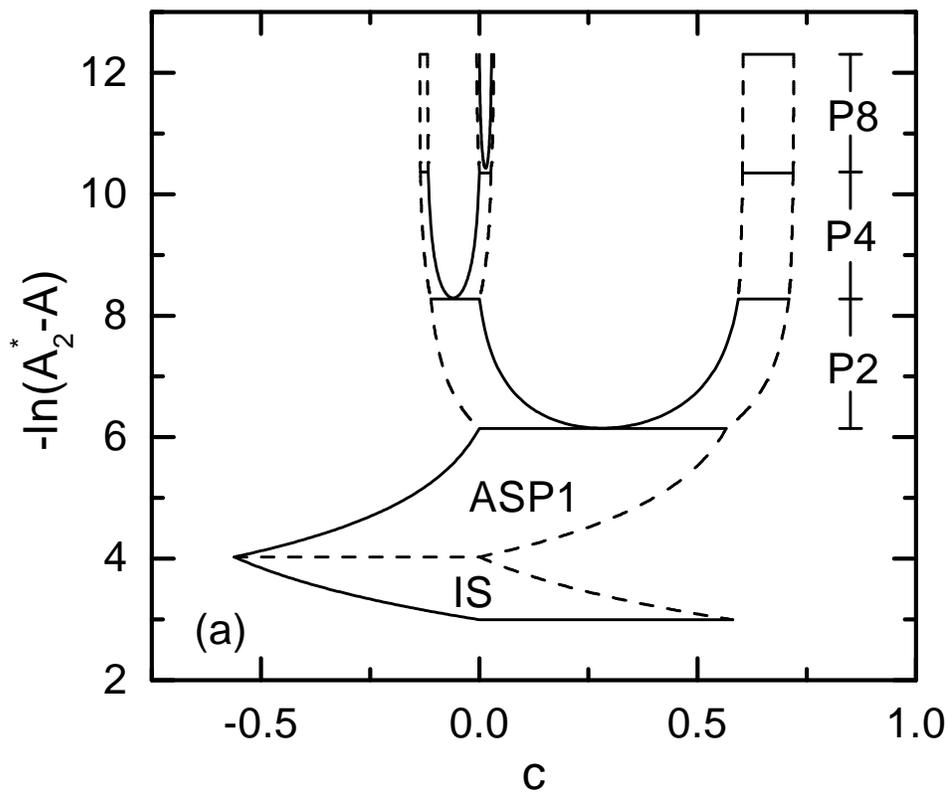
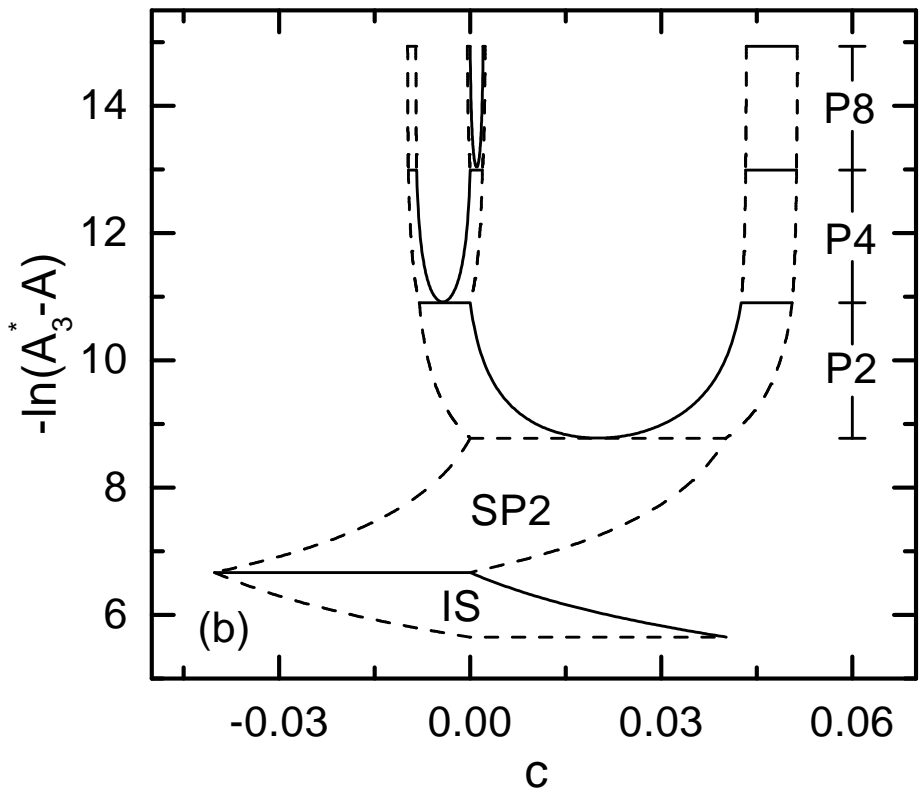
\begin{figure}
\caption{(a) Second and (b) third stability diagrams of synchronous
    periodic orbits. Here $A^*_2$ $(=3.829\,784\, \cdots)$ and
    $A^*_3$ $(=10.675\,090)$ are just the second and third 
    period-doubling transition points of the uncoupled inverted
    pendulum, respectively. The stable regions of the inverted 
    stationary point, an asymmetric orbit of period $1$, a 
    symmetric $2$-periodic orbit, and an asymmetric N-periodic 
    $(N=2,4,8)$ orbit are denoted by the IS, the ASP1, the SP2 and 
    the PN, respectively. The solid and dashed boundary lines 
    also represent the same as those in Fig.~\ref{SD1}. 
     }
\label{SD3}
\end{figure}

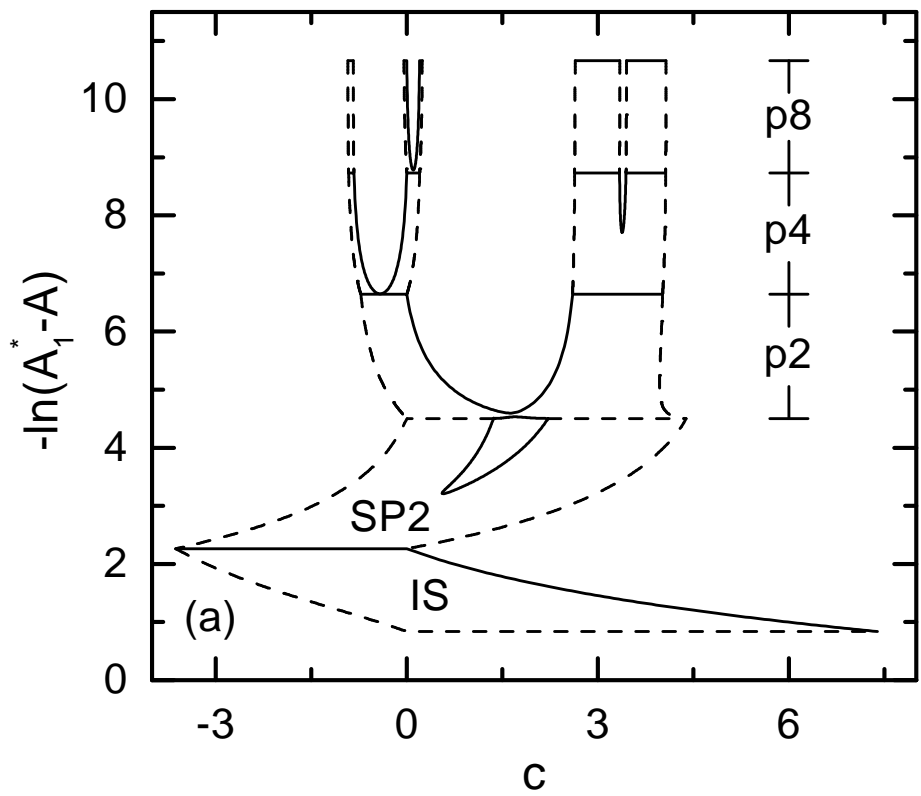
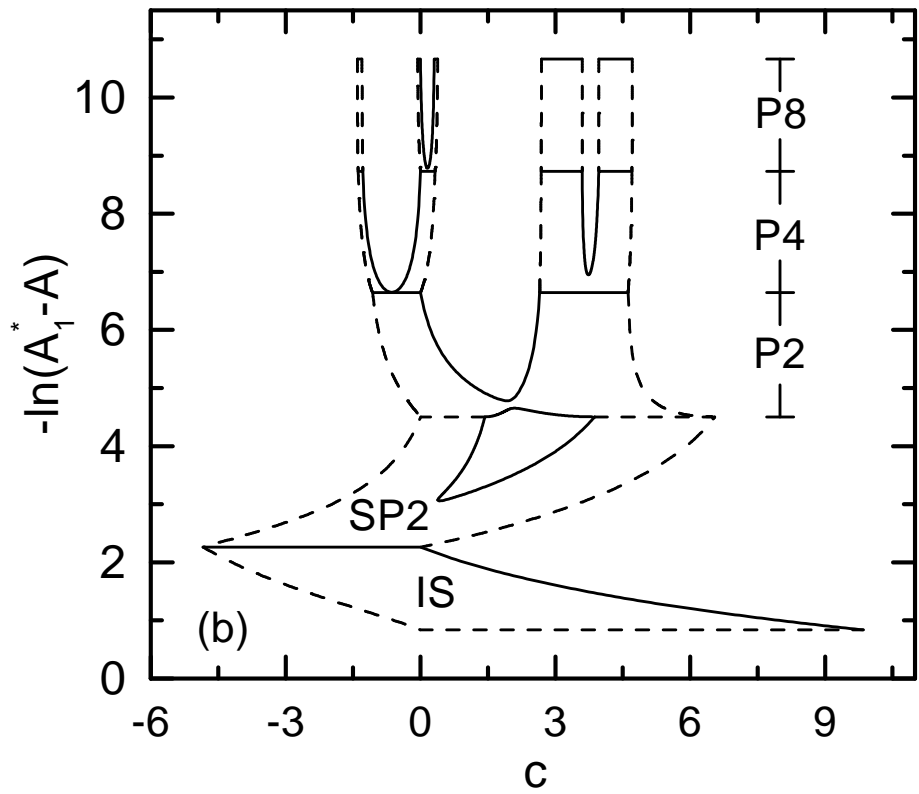
\begin{figure}
\caption{ First stability diagrams of synchronous periodic orbits near
the $c=0$ line for the cases of (a) the quadratic and (b) cubic
couplings. Here SP2 and PN (N$=2,4,8$) denote the stable regions of
a symmetric orbit of period $2$ and an asymmetric orbit with period N,
respectively.
         }
\label{NCSD}
\end{figure}

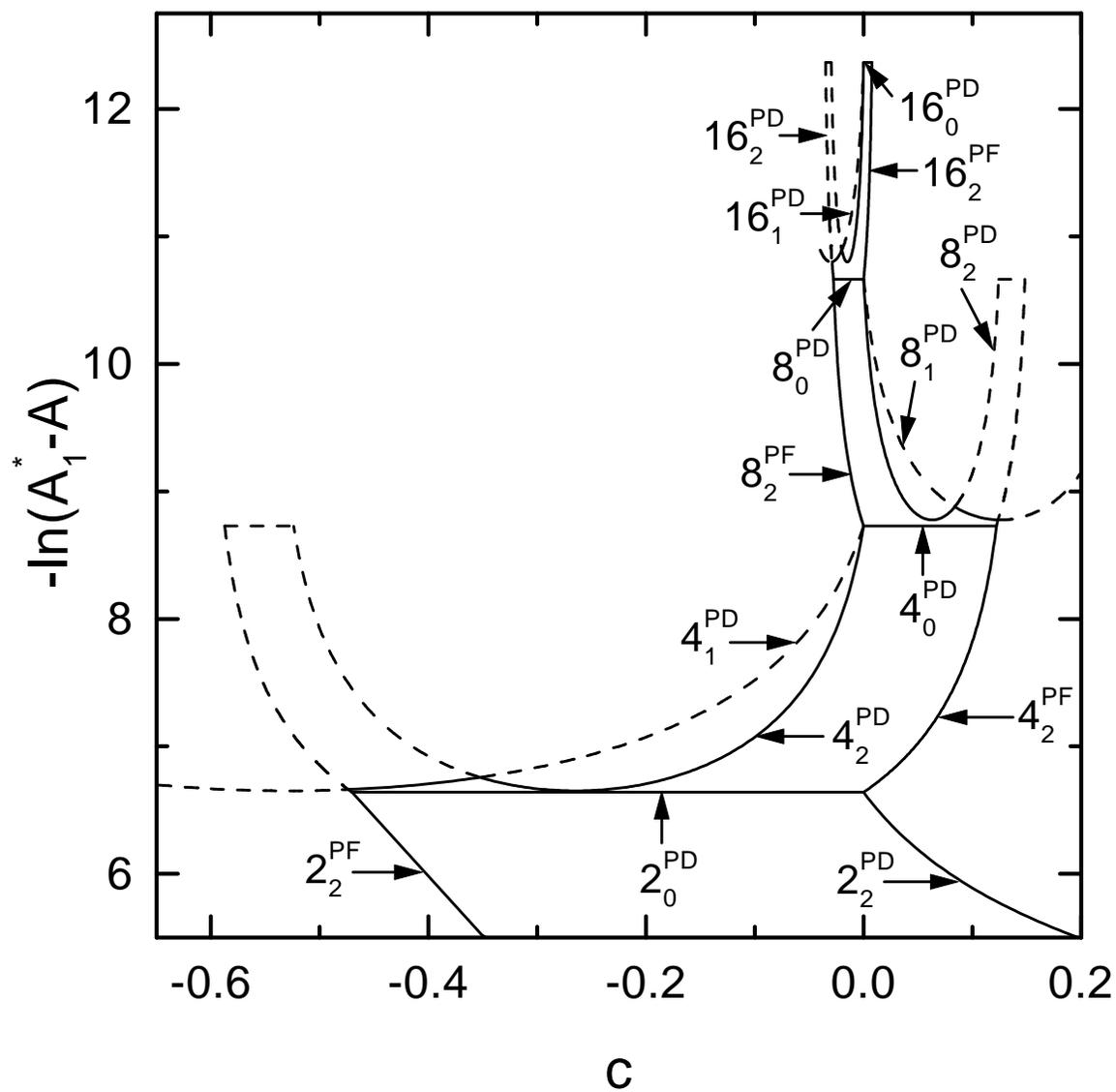
\begin{figure}
\caption{ Stability diagram of synchronous periodic orbits in four
linearly-coupled inverted pendulums with the nearest-neighbor coupling
($K=1$). Each stable region is bounded by its solid boundary
curves. For a synchronous orbit of period $q$, the PDB (PFB) curve
of the mode with index $j$ is denoted by a symbol $q^{PD(PF)}_j$.
     }
\label{MCSD}
\end{figure}


\begin{references}
\bibitem{Rem} This is also valid for the case of the level $1$.
That is, in the case of level $1$ the zero-$c$ left side of
$U_4(1,2)$ remains unchanged, whereas the other right side becomes
flattened by the bifurcation curve of the asynchronous mode with
$j=1$. Consequently, the chimney (consisting of an infinite pile
of rectangular-shape regions) in Fig.~\ref{SD2}(b) disappears for
the nonglobal-coupling case.
\end{references}
\end{document}